# Observation of superconductivity in 3D Dirac semimetal $Cd_3As_2$ crystal

He Wang[1,2]†, Huichao Wang[1,2]†, Haiwen Liu[1,2]†, Hong Lu[1,2], Wuhao Yang[1,2], Shuang Jia[1,2], Xiong-Jun Liu[1,2]*, X. C. Xie[1,2], Jian Wei[1,2]*, and Jian Wang[1,2]*

[1]*International Center for Quantum Materials, School of Physics, Peking University, Beijing 100871, China*

[2]*Collaborative Innovation Center of Quantum Matter, Beijing, China*

†These authors contributed equally to this work.

*e-mail: jianwangphysics@pku.edu.cn; weijian6791@pku.edu.cn; xiongjunliu@pku.edu.cn

**Lately, the three-dimensional (3D) Dirac semimetal, which possesses 3D linear dispersion in electronic structure as a bulk analogue of graphene, has generated widespread interests in both material science and condensed matter physics[1,2]. Very recently, crystalline $Cd_3As_2$ has been proposed and proved to be one of 3D Dirac semimetals which can survive in atmosphere[3-9]. Here, by controlled point contact (PC) measurement, we observe the exotic superconductivity around point contact region on the surface of $Cd_3As_2$ crystal. The observation of zero bias conductance peak (ZBCP) and double conductance peaks (DCPs) symmetric to zero bias further reveal p-wave like unconventional superconductivity in $Cd_3As_2$ quantum matter. Considering the topological property of the 3D Dirac semimetal, our findings may indicate that the $Cd_3As_2$ crystal under certain conditions is a candidate of the topological superconductor[10-13], which is predicted to support Majorana zero modes or gapless Majorana edge/surface modes in the boundary depending on the dimensionality of the material[14-17].**

The 3D Dirac semimetal state is located in the topological phase boundary and can potentially be driven into other topological phases including topological insulator, topological metal, Weyl semimetal and even topological superconductor states. The realization of topological superconductivity and Majorana fermions is of great interest due to both the exploration of fundamental physics and the potential applications in fault-tolerant topological quantum computation[18,19]. Nevertheless, so far the experimental demonstration is still under debate despite significant progresses having been achieved in this direction[20-28]. The discovery of new and reliable topological superconductor candidates is strongly called for.

In our experiment, single crystal of $Cd_3As_2$ is synthesized from a Cd-rich melt with stoichiometry $Cd_8As_2$ in the evacuated quartz ampoule[9]. The observed unusual physical properties such as ultrahigh mobility, low carrier density, huge linear magnetoresistance and the sophisticated geometry of Fermi surface have made $Cd_3As_2$ crystal a promising quantum material[6-9]. In this paper, the transport results are mainly from two typical $Cd_3As_2$ samples (S1, S2) from different batches. As shown in Fig.1a, the standard four-electrode measurement demonstrates the metallic like behavior of S1, which is consistent with previous reports[7-9]. Besides, the non-superconducting behavior down to 2 K is further confirmed by magnetization measurement for the same sample (See Supplementary Information). Point contact spectroscopy (PCS) measurements are conducted with the most common "needle-anvil" configuration[29]. Tungsten wire of 0.25 mm diameter is etched to form a sharp tip, and is fixed on top of the (112) surface plane of the $Cd_3As_2$ single crystal sample mounted on an Attocube nanopositioner stack (the inset of Fig. 1b). Since the tip and sample are both fixed to the same housing, which is suspended with springs at the bottom of the probe of a Leiden dilution fridge, relative vibration and displacement between the tip and sample is small during field and temperature ramping, ensuring a stable contact and repeatable

PCS. One advantage of our customized setup is that we can measure PCS at different locations on the surface of the same sample in one run, and this helps to confirm that the measured PCS is intrinsic to the sample surface. Differential resistance (dV/dI) spectra are measured by standard lock-in technique in quasi-four-probe configuration. Figure 1b shows the PC transport results between tungsten tip and $Cd_3As_2$ single crystal. A significant resistance drop is observed in the process of zero field cooling down. The onset temperature ($T_{onset}$) of this resistance drop is 3.9 K, which is considered to be a signature of superconductivity since resistance vs. temperature measurement in a perpendicular magnetic field of 625 Oe shows $T_{onset}$ decreases to 3.7 K (Fig. 1b). 'Soft' PC measurement[29] is also performed but no superconducting behavior is observed (See Supplementary Information). Thus, the observed superconductivity is likely induced by the pressure of the tip contact, rather than an intrinsic property of $Cd_3As_2$.

A set of point contact spectra (normalized dI/dV vs. bias voltage) of 5 Ω PC resistance at different temperatures is shown in Fig.1c. The PC spectrum of 0.28 K is marked with bold blue line, in which DCPs can be observed at ± 0.9 meV and disappear as the temperature is increased above 2.0 K. Double conductance dips at 0.28 K locate at ± 1.5 meV symmetrically around the zero bias and disappear for temperatures above 3.6 K. We also notice that there is a small ZBCP in the PCS at lower temperatures than 2.8 K (Figs. 1c and d). The presence of ZBCP in PCS could be due to unconventional superconductivity. The observed DCPs and double conductance dips features, which exist independently of the ZBCP (See Supplementary Information), cannot be fitted with the conventional Blonder-Tinkham-Klapwijk (BTK) theory[30]. Similar DCP features were investigated for the possible p-wave superconductor $Sr_2RuO_4$ using point contact[31] and in-plane tunnel junctions[32]. On the other hand, the ZBCP may relate to the Majorana zero modes which are predicted to exist in the ends of a one-dimensional (1D) topological superconductor or the vortex cores of a two-dimensional (2D) topological superconductor[12].

The observed superconductivity behavior is further confirmed by another 7.5 Ω PCS measurement at a different position on the same surface of S1. An onset PC resistance drop temperature of 3.9 K is apparent for this point contact too (inset of Fig. 2a). The PCS under different perpendicular magnetic field has also been examined at 3.8 K to identify the superconductivity feature (Fig.2a). We can see an obvious suppression of the superconducting conductance peak around zero bias voltage with the applied perpendicular magnetic field increasing from 0 Oe to 625 Oe. The temperature dependence of the PCS is shown in Fig. 2b. Similar DCP features locating at ±1 meV is also observed at 0.28 K, which disappear as the temperature increases above 2.5 K. The weak ZBCP is also observed below 0.8 K (inset of Fig. 2b).

To further demonstrate our observation of superconductivity in 3D Dirac semimetal $Cd_3As_2$, the PCS of another sample (S2) was studied. Figure 3 shows the temperature dependence of the PC resistance and a set of normalized PCS at different temperatures. A $T_{onset}$ of 7.1 K is observed in this sample (Fig. 3a) indicating a critical temperature $T_C$ up to 7.1 K. A wide zero bias conductance hump can be clearly observed in the spectra (Fig. 3b), which gradually disappears with increasing temperature. Complete suppression of the conductance hump occurs around 7 K, corresponding to the onset temperature of 7.1 K and representing the superconducting behavior. As the temperature is decreased below 1.2 K, a ZBCP as well as DCPs appears (Fig. 3c), qualitatively consistent with our observation in S1 (Fig.1d). The ZBCP becomes sharper with decreasing temperature.

The main features of dI/dV spectra can be classified into two types: (i) a ZBCP observed at low temperature; (ii) DCPs and double conductance dips both symmetric to zero bias, with the peaks smearing into a broad hump at high temperature. These exotic spectra features can be attributed to the possible Majorana modes of the superconducting $Cd_3As_2$. The system may exhibit profound topological superconducting phases in the surface and bulk, as summarized into the schematic phase diagram shown in Fig. 4a. For the sake of brevity, we firstly give out the phase diagram along $k_z$ [001], and the phase diagram along [112] can be resolved by projection of the $k_z$ [001] onto the [112] direction. The phase diagram consists of three regions which are separated by gapless nodal points: quasi-1D chiral orthogonal[13] or BDI class topological superconductor (BDI-SC) (region I) obtained in the surface, quasi-2D helical p-wave superconductor (HPSC) in the bulk coexisting with normal s-wave superconductor (NSC) ( region II) and normal insulator (NI) ( region III). For each fixed $k_{z3}$ and -$k_{z3}$ in region I, the $Cd_3As_2$ system without superconductivity is captured by a quasi-2D quantum spin Hall insulator (QSHI) with helical edge states, which are Fermi arc states (FAS) of the Dirac semimetal crossing the Fermi energy (see Fig. 4b and 4e). The presence of an s-wave pairing opens a gap in the helical edge states (FAS) which renders a quasi-1D BDI-SC with Majorana zero modes localized in the boundary of the superconducting area (see Fig. 4f)[12]. On the other hand, in region II, both intra-valley and inter-valley pairings can exist in the bulk states of the Dirac semimetal (see Fig. 4b). Specifically, the intra-valley paring between slice $k_{z1}$ and slice $k_{z2}$ in this region occurs between two states belonging to a quasi-2D QSHI and quasi-2D NI (see Fig. 4c), respectively. The resulted effective Hamiltonian resembles a quasi-2D HPSC with helical Majorana edge modes. The total number of Majorana zero modes corresponding to region I and branches of helical Majorana edge modes for region II is determined by the number of momenta $k_z$ taken in the two regions, and all these Majorana modes contribute to the dI/dV spectra. The phase boundary between quasi-2D HPSC (region II) and quasi-1D BDI-SC (region I) or NI (region III) is $\Delta^2 + E_F^2 = 4M_1^2 K^2 (\delta k_z)^2$, here $\delta k_z$ denotes the mometum along [001] direction measured from the Weyl point $K$, and $M_1$ denotes certain parameter of $Cd_3As_2$ (See Supplementary Information). Meanwhile, for the inter-valley paring between slices $k_{z1}$ (QSHI) and $-k_{z1}$ (QSHI), the effecive pairing gives a NSC (See Supplementary Information).

In order to compare with the experimental observation of (112) surface, we project the above phase diagram onto the one versus the [112] momentum. Since the projection covers all the FAS, the Majorana zero modes of the quasi-1D BDI-SCs for each paired $k_{z3}$ channel contribute to the ZBCP. For the quasi-2D HPSCs, the total number of Majorana edge modes of energy $E$ in region II is maximized at $E=0$ and decreases with increasing energy, implying that these Majorana zero edge modes also contribute to ZBCP. Note that the projection onto the [112] direction can deform the quasi-2D HPSCs to be anisotropic due to the anisotropy of Ferimi surface[2]. Thus the helical Majorana edge modes with finte energy may account for the DCPs in the dI/dV spectra by examing the angle-resolved conductance spectra with an anisotropic pairing[32]. Moreover, the double conductance dips in dI/dV spectra may be attributed to the quasi-2D HPSCs with the extended BTK theory taking into account of order parameter symmetries[32]. Note that the DCPs might also correspond to the quasiparticle gap of the NSC, and the double conductance dips may possibley explained by the critical current effect due to the heating in the contact region[33]. However, due to the high mobility of single crystal $Cd_3As_2$, the PC is close to the ballistic limit with the Sharvin resistance $R_S$ much larger than Maxwell resistance $R_M$, which is not well consistent with the critical current effect scenario.

In summary, by point contact measurement using "needle-anvil" configuration we observe the unexpected superconducting behavior in 3D Dirac semimetal $Cd_3As_2$ crystal. The observed ZBCP and DCPs features in PCS indicate the superconductivity we found is unconventional. Our further theoretical analyses reveal that the exotic features may originate from topological superconductivity in the surface FAS and bulk states. In particular, the surface of the superconducting Dirac semimetal renders quasi-1D BDI-SCs which support Majorana zero modes associated with the ZBCP, while the bulk can exhibit as the quasi-2D HPSCs which support helical Majorana edge modes that may contribute to both the ZBCP and DCPs. So far, most experiments are still focused on demonstrating or exploring the existence of topological superconductors. Our discovery of superconductivity in 3D Dirac semimetal which is of ultrahigh mobility may offer an ideal candidate or platform for studying topological superconductivity.

Note added: We noted that another group also studied PCS of polycrystalline $Cd_3As_2$ and showed an indication of superconductivity in this compound[34].

## Acknowledgements

We acknowledge Chenglong Zhang, Fan Yang, Ying Xing and Yi Liu for the help in experiments. This work was financially supported by National Basic Research Program of China (Grant Nos. 2013CB934600, 2012CB921300, 2012CB927400), the National Natural Science Foundation of China (Nos. 11222434, 11174007), and the Research Fund for the Doctoral Program of Higher Education (RFDP) of China.

## Author Contributions

J.W. and J.W. conceived the experiments. H.W., H.W. and W.Y. carried out transport measurements. H.L., X.J.L. and X.C.X. performed the theoretical interpretation. H.L. and S.J. grew the crystals.

## Additional information

Supplementary information is available in the online version of the paper. Reprints and permissions information is available online at www.nature.com/reprints.
Correspondence and requests for materials should be addressed to J.W., J.W. and X.J.L.

## Competing Financial Interests statement

The authors declare no competing financial interests.

# Figure Legends

Figure 1| **Transport measurements of $Cd_3As_2$ (S1). a**, Temperature dependence of four-probe bulk resistivity showing typical metallic behavior with a residual resistance at low temperatures. Upper Inset: Schematics of the standard four-probe method measurement configuration. Lower inset: Zoom-in of the Resistance-Temperature curve below 10 K. **b**, Temperature dependence of the zero bias PC resistance. The onset critical temperature decreases from 3.9 K to 3.7 K when the applied out-of-plane magnetic field is 625 Oe. Inset: Schematics of the PCS measurement configuration. **c**, The normalized dI/dV spectra at different temperatures varying from 0.28 K to 3.8 K without external magnetic field. **d**, Zoom-in of the normalized ZBCP.

Figure 2| **Superconducting features in S1 at a point contact resistance of 7.5 Ω. a**, The normalized dI/dV spectra at 3.8 K under different out-of-plane magnetic fields. Inset: Temperature dependence of the PC resistance showing clear superconducting transition. **b**, The normalized dI/dV spectra for the temperature range from 0.28 to 4.0 K. Inset: ZBCP at selected temperatures.

Figure 3| **Transport results of another $Cd_3As_2$ sample (S2). a.** The PC resistance vs. temperature curve of S2, showing an onset Tc of 7.1 K. Inset: Temperature dependence of the bulk resistivity measured by standard four-probe method, showing non-superconducting behavior. **b.** Normalized PCS at a PC resistance of 65 Ω for the temperature from 0.28 K to 4.0 K. **c**, ZBCP at different temperatures.

**Figure 4│ Topological phase diagram for superconducting Cd₃As₂. a,** The phase diagram consists of three regions: quasi-1D BDI-SC, quasi-2D HPSC coexisting with NSC, and NI. **b,** The Fermi surface with different $k_z$ slices. **c,** The schematic of intra-valley pairing. **d,** The schematic of inter-valley pairing. **e-f,** The schematic of pairing between FAS, which behaves as quasi-1D BDI-SC with Majorana zero modes.

# Figure 1

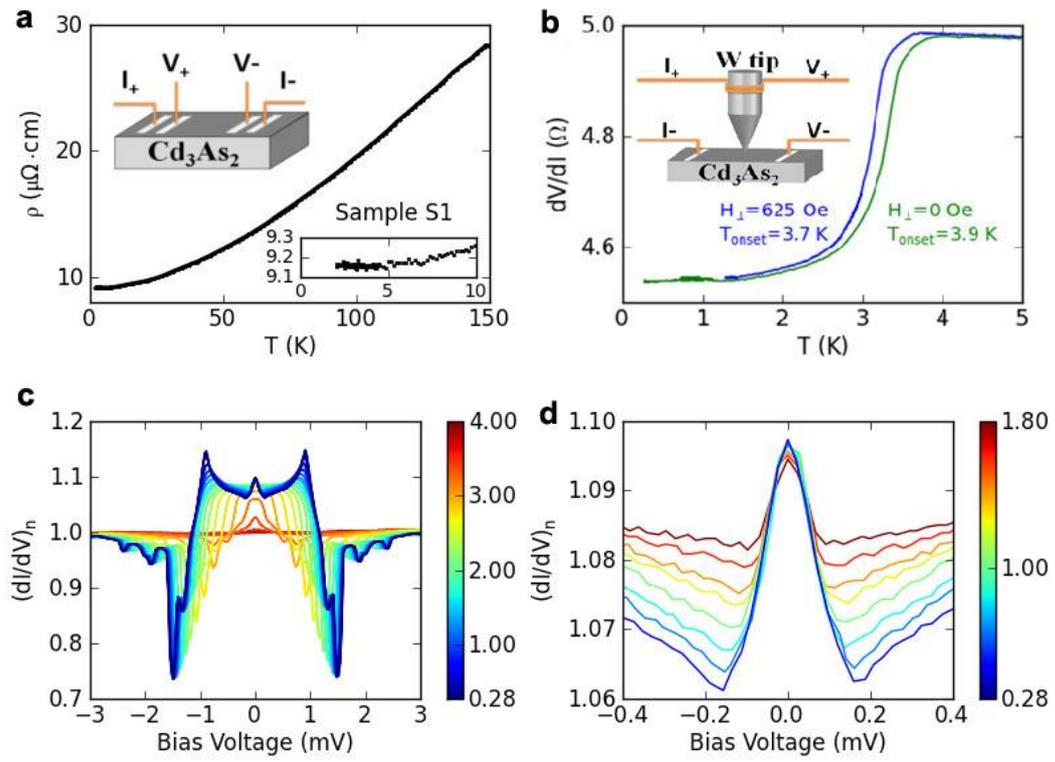

# Figure 2

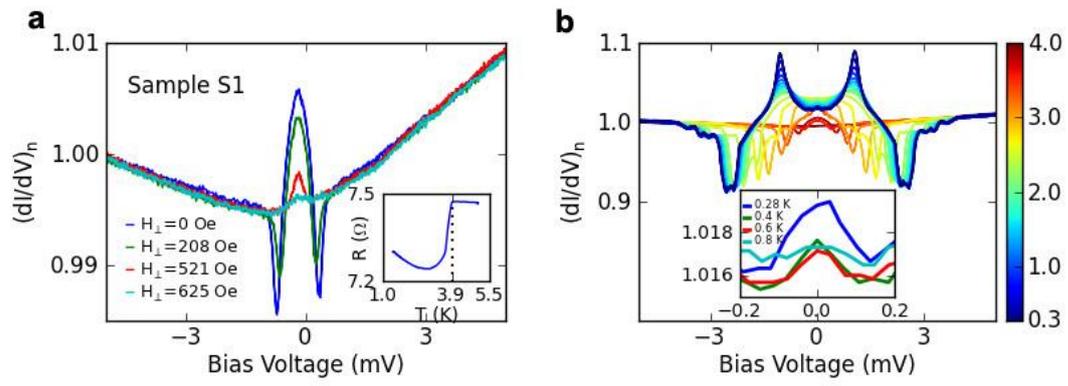

**Figure 3**

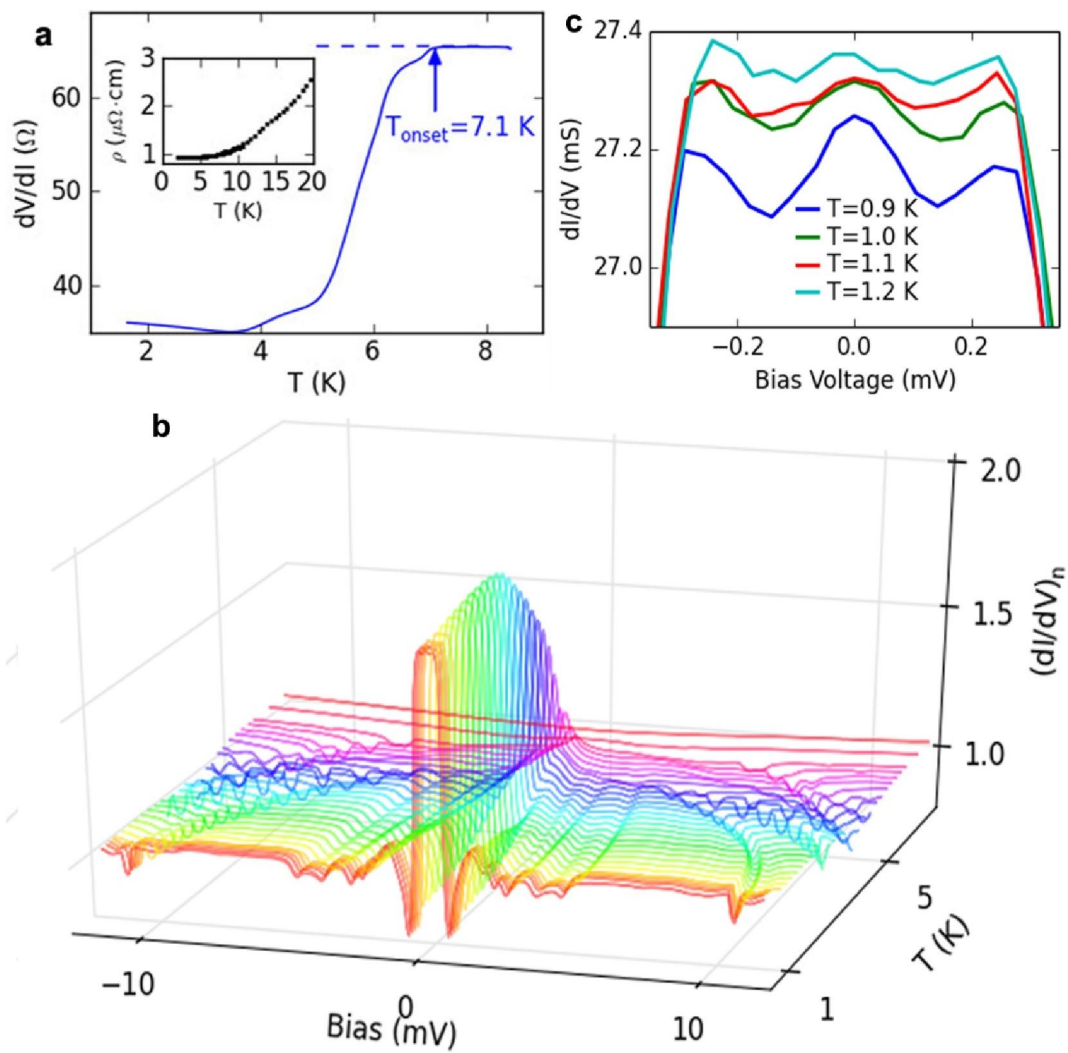

**Figure 4**

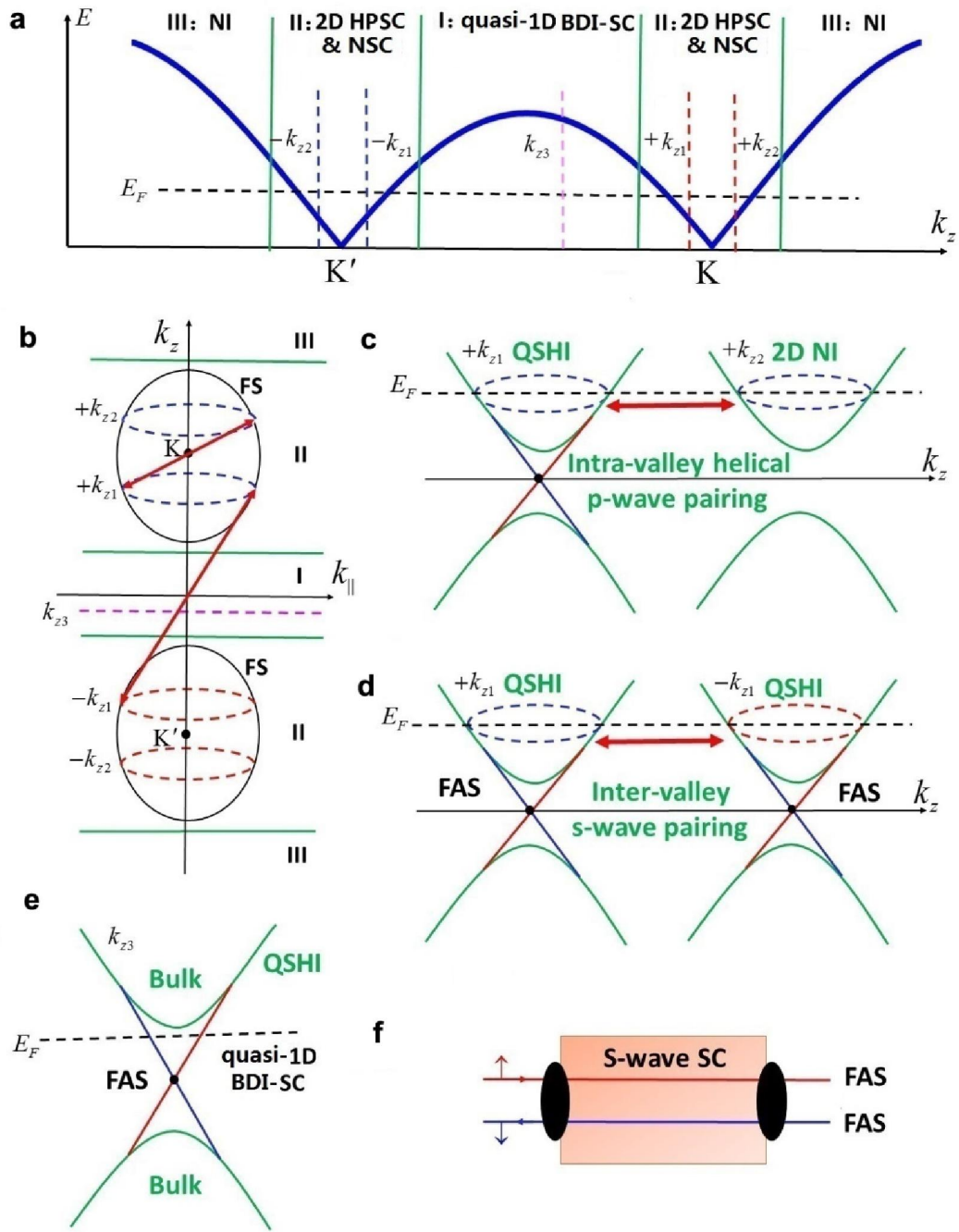

# Supplementary Information
# Observation of superconductivity in 3D Dirac semimetal $Cd_3As_2$ crystal

He Wang[1,2]†, Huichao Wang[1,2]†, Haiwen Liu[1,2]†, Hong Lu[1,2], Wuhao Yang[1,2], Shuang Jia[1,2], Xiong-Jun Liu[1,2]*, X. C. Xie[1,2], Jian Wei[1,2]*, and Jian Wang[1,2]*

[1]*International Center for Quantum Materials, School of Physics, Peking University, Beijing 100871, China*

[2]*Collaborative Innovation Center of Quantum Matter, Beijing, China*

†These authors contributed equally to this work.

*e-mail: jianwangphysics@pku.edu.cn; weijian6791@pku.edu.cn; xiongjunliu@pku.edu.cn

I. **Magnetization measurements of the $Cd_3As_2$ crystal.**

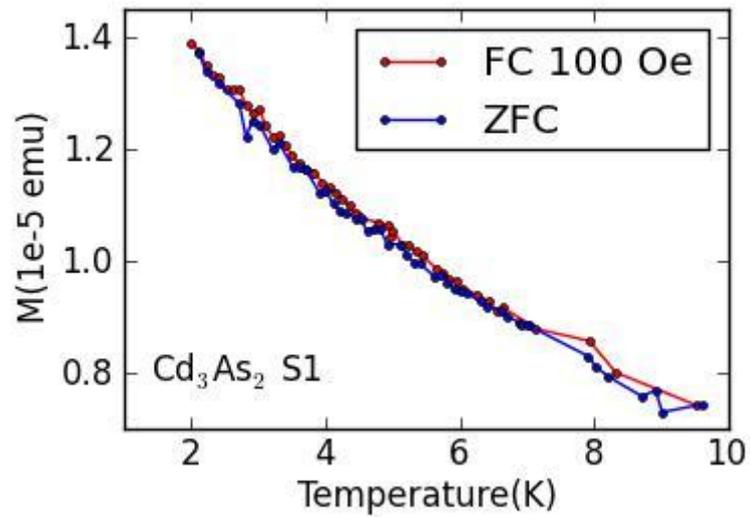

**Figure S1 │Magnetization property of the $Cd_3As_2$ crystal.** The Magnetization vs. temperature (M-T) curves of sample 1 (S1) during both zero field cooling (ZFC) and field cooling (FC) at 100 Oe. The magnetic field is perpendicular to the (112) surface plane. No superconducting feature is observed down to 2 K. The measurements were done in a Quantum Design Magnetic Property Measurement System (MPMS XL-17).

## II.  'Soft' point contact measurements of the $Cd_3As_2$ crystal.

The so-called 'soft' point contact[1] is formed by a tiny drop of silver paint on the clean surface of $Cd_3As_2$ crystal connecting a 25-μm-diameter gold wire. The soft point contacts are prepared at room temperature in ambient atmosphere. No superconducting behavior is observed by this method. The results from two typical $Cd_3As_2$ samples are shown in Fig. S2.

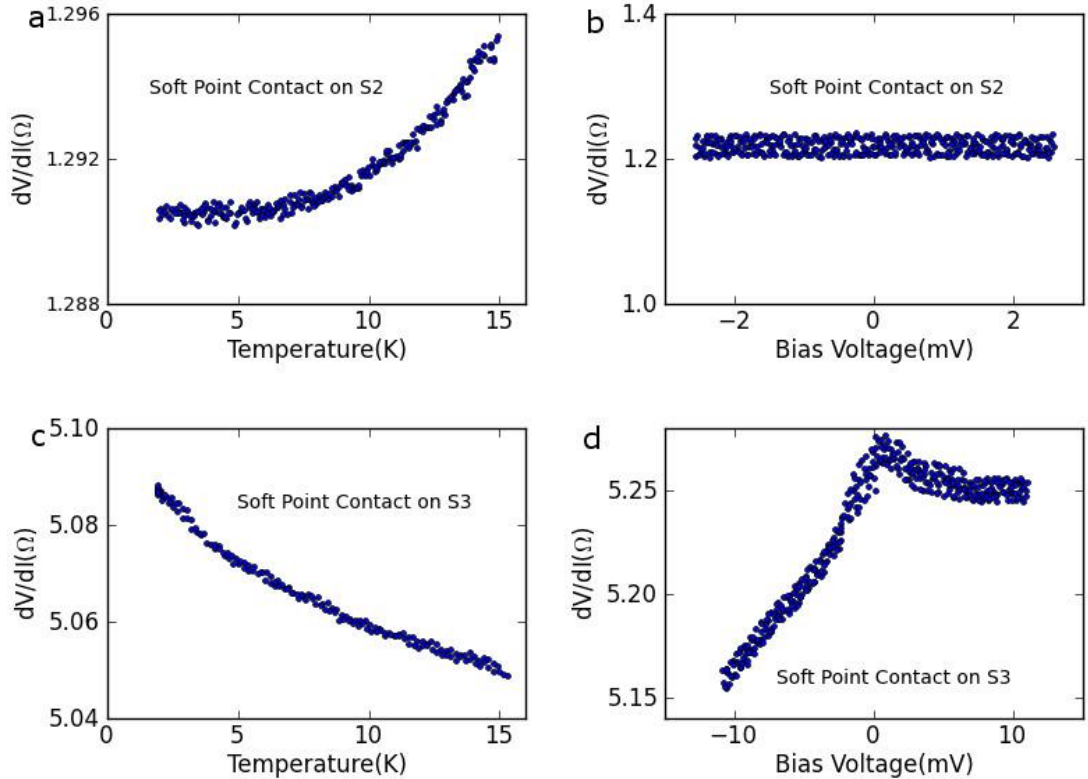

**Figure S2 │Non-superconducting features of the $Cd_3As_2$ crystal measured by the 'soft' point contact technique. a**. Temperature dependence of the zero bias 'soft' point contact resistance of sample 2 (S2). **b**. The dV/dI spectrum of S2 at 2 K without external magnetic field. **c**. The zero bias 'soft' point contact resistance of sample 3 (S3) as a function of temperature. **d**. The dV/dI spectrum of S3 at 2 K at zero magnetic field.

# III. Independent existence of zero bias conductance peak and double conductance peaks.

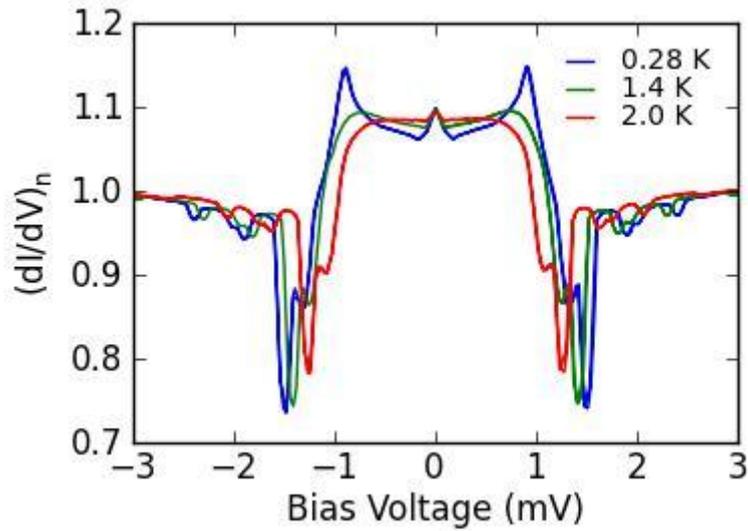

**Figure S3 │ Zero bias conductance peak (ZBCP) and double conductance peaks (DCPs) at selected temperatures.** Both ZBCP and DCPs are observed at 0.28 K. When the temperature is increased, the DCPs is evolved a hump and the ZBCP becomes broader. At relatively high temperature 2.0 K, the ZBCP survives while the DCPs disappear, indicating the independent existence of them.

## IV. Detailed derivation for the 2D helical p-wave pairing.

Considering the basis set of $\left|S_{1/2},\uparrow\right\rangle$, $\left|P_{3/2},\uparrow\right\rangle$, $\left|S_{1/2},\downarrow\right\rangle$ and $\left|P_{3/2},\downarrow\right\rangle$ states, the effective low energy Hamiltonian near $\Gamma$ pointe is given by[2]:

$$H_\Gamma(\vec{k}) = \varepsilon_0(\vec{k}) + \begin{pmatrix} M(\vec{k}) & Ak_+ & Dk_- & B^*(\vec{k}) \\ Ak_- & -M(\vec{k}) & B^*(\vec{k}) & 0 \\ Dk_- & B(\vec{k}) & M(\vec{k}) & -Ak_- \\ B(\vec{k}) & 0 & -Ak_+ & -M(\vec{k}) \end{pmatrix} \quad \text{(SE1)},$$

where $k_\pm = k_x \pm ik_y$, and $M(\vec{k}) = M_0 - M_1 k_z^2 - M_2(k_x^2 + k_y^2)$ with $M_0$, $M_1$, $M_2 < 0$. For simplicity, we neglect the $\varepsilon_0(\vec{k})$, $D$ and $B^*(\vec{k})$ terms. The Weyl points $(0,0,\pm k_{zc})$ satisfies the relation $M_0 - M_1 k_{zc}^2 = 0$. Now expanding the Hamiltonian near the slices $k_{z1} = k_{zc} - \delta k_z$, $k_{z2} = k_{zc} + \delta k_z$, $-k_{z1} = -k_{zc} + \delta k_z$ and $-k_{z2} = -k_{zc} - \delta k_z$ (See Fig. 4b), we have the 2D effective Hamiltonian for certain $k_z$ slice:

$$H(\pm k_{z1}, k_\parallel) = \begin{pmatrix} M(\delta k_z) - M_2 k_\parallel^2 & Ak_+ & 0 & 0 \\ Ak_- & -M(\delta k_z) + M_2 k_\parallel^2 & 0 & 0 \\ 0 & 0 & M(\delta k_z) - M_2 k_\parallel^2 & -Ak_- \\ 0 & 0 & -Ak_+ & -M(\delta k_z) + M_2 k_\parallel^2 \end{pmatrix}$$

(SE2).

$$H(\pm k_{z2}, k_\parallel) = \begin{pmatrix} -M(\delta k_z) - M_2 k_\parallel^2 & Ak_+ & 0 & 0 \\ Ak_- & M(\delta k_z) + M_2 k_\parallel^2 & 0 & 0 \\ 0 & 0 & -M(\delta k_z) - M_2 k_\parallel^2 & -Ak_- \\ 0 & 0 & -Ak_+ & M(\delta k_z) + M_2 k_\parallel^2 \end{pmatrix}$$

(SE3).

, here $M(\delta k_z) = 2M_1 k_{zc} \cdot \delta k_z < 0$. Thus, it is very clear that $H(\pm k_{z1}, k_\parallel)$ are 2D QSHI and $H(\pm k_{z2}, k_\parallel)$ are 2D normal insulator. Next, we consider the superconductor pairing between the Fermi surfaces. For intra-valley pairing between $H(k_{z1},k_{\parallel})$ and $H(k_{z2},k_{\parallel})$, we give out the detailed BdG Hamiltonian for $|S,\uparrow\rangle_{k_{z1},k_{\parallel}}$, $|P,\uparrow\rangle_{k_{z1},k_{\parallel}}$, $|S,\downarrow\rangle^{\dagger}_{k_{z2},-k_{\parallel}}$ and $|P,\downarrow\rangle^{\dagger}_{k_{z2},-k_{\parallel}}$:

$$H_{BdG} = \begin{pmatrix} M(\delta k_z) - M_2 k_{\parallel}^2 - E_F & Ak_+ & \Delta & 0 \\ Ak_- & -M(\delta k_z) + M_2 k_{\parallel}^2 - E_F & 0 & \Delta \\ \Delta & 0 & M(\delta k_z) + M_2 k_{\parallel}^2 + E_F & -Ak_+ \\ 0 & \Delta & -Ak_- & -M(\delta k_z) - M_2 k_{\parallel}^2 + E_F \end{pmatrix}$$

(SE4).

The gap close condition is given by: $\Delta^2 + E_F^2 = [M(\delta k_z)]^2$. The region II in Fig. 4a satisfies $\Delta^2 + E_F^2 > [M(\delta k_z)]^2$, which is connected to a reprehensive simple case with $E_F = 0$ and $M(\delta k_z) = 0$, namely the $k_z$ slice penetrates the Weyl point. Thus, the BdG Hamiltonian further simplifies to a chiral superconductor with Chern number $N = -1$:

$$H_{BdG} = \begin{pmatrix} -M_2 k_{\parallel}^2 & Ak_+ & \Delta & 0 \\ Ak_- & M_2 k_{\parallel}^2 & 0 & \Delta \\ \Delta & 0 & M_2 k_{\parallel}^2 & -Ak_+ \\ 0 & \Delta & -Ak_- & -M_2 k_{\parallel}^2 \end{pmatrix}$$ (SE5).

The other four time reversal basis in Nambu representaion $|S,\downarrow\rangle^{\dagger}_{k_{z1},-k_{\parallel}}$, $|P,\downarrow\rangle^{\dagger}_{k_{z1},-k_{\parallel}}$, $|S,\uparrow\rangle_{k_{z2},k_{\parallel}}$ and $|P,\uparrow\rangle_{k_{z2},k_{\parallel}}$ results in a chiral superconductor with Chern number $N = 1$. In total, intra-valley pairing between $H(k_{z1},k_{\parallel})$ and $H(k_{z2},k_{\parallel})$ results in a 2D helical p-wave superconductor with helical majorana edge modes. And the phase boundary condition between 2D 2D HPSC and other phases is given by $\Delta^2 + E_F^2 = [M(\delta k_z)]^2$. Similar as the intra-valley case, we check the inter-valley pairing between $H(k_{z1},k_{\parallel})$ and $H(-k_{z1},k_{\parallel})$, and we find a normal superconductor without topological expiations.

Similarly, the 2D effective Hamiltonian for the $k_{z3}$ slice reads:

$$H(k_{z3}, k_\parallel) = \begin{pmatrix} M(\delta k_{z3}) - M_2 k_\parallel^2 & Ak_+ & 0 & 0 \\ Ak_- & -M(\delta k_{z3}) + M_2 k_\parallel^2 & 0 & 0 \\ 0 & 0 & M(\delta k_{z3}) - M_2 k_\parallel^2 & -Ak_- \\ 0 & 0 & -Ak_+ & -M(\delta k_{z3}) + M_2 k_\parallel^2 \end{pmatrix}$$

(SE6),

where $\delta k_{z3} = k_{zc} - k_{z3} > 0$, and $M(\delta k_{z3}) = 2M_1 k_{zc} \cdot \delta k_{z3} < 0$. Thus, the $M(\delta k_{z3}) - M_2 k_\parallel^2$ change sign in the $k_\parallel$ plane, and $H(k_{z3}, k_\parallel)$ corresponds to a QSHI. As shown in Figure 4 in main text, the Fermi surface penetrates through the helical edge states. Thus, the effective 1D Hamiltonian for edge states is: $H_{1D} = \hbar v_0 \sigma_z k_1 - E_F$, with $v_0$, $\sigma_z$ and $k_1$ denotes the Fermi velocity, spin and momentum of the helical edge states, respectively. When s-wave pairing is tuned on between edge states with opposite spin, the system with a fixed $k_z$ becomes a quasi-1D topological superconductor which is equivalent to Kitaev's model with Majorana zero modes in the boundary of the superconducting area[3]. Each $k_z$ slice in region I of Fig. 4a possesses two Majorana zero modes, and contribute to the dI/dV spectra.